\documentclass[reprint,prx,amsmath,amssymb,longbibliography,superscriptaddress]{revtex4-1}

\pdfoutput=1

\usepackage{graphicx}
\usepackage{amsmath,latexsym,tabularx}
\usepackage{xcolor}
\usepackage{multirow}
\usepackage[
  colorlinks=true,
  urlcolor=blue,
  linkcolor=blue,
  citecolor=blue
]{hyperref}
\usepackage{multirow,booktabs}

\newcommand{\affilLL}[0]{Lincoln Laboratory, Massachusetts Institute of Technology, Lexington, Massachusetts 02421, USA}
\newcommand{\affilMIT}{Massachusetts Institute of Technology, Cambridge, Massachusetts 02139, USA}

\newcommand{\ca}{Ca$^{+}$}
\newcommand{\sr}{Sr$^{+}$}

\begin{document}

\title{Dual-Species, Multi-Qubit Logic Primitives for Ca$^{+}$/Sr$^{+}$ Trapped-Ion Crystals}

\author{C. D. Bruzewicz}
\email[]{colin.bruzewicz@ll.mit.edu}
\affiliation{\affilLL}

\author{R. McConnell}
\affiliation{\affilLL}

\author{J. Stuart}
\affiliation{\affilLL}
\affiliation{\affilMIT}

\author{J. M. Sage}
\email[]{jsage@ll.mit.edu}
\affiliation{\affilLL}
\affiliation{\affilMIT}

\author{J. Chiaverini}
\email[]{john.chiaverini@ll.mit.edu}
\affiliation{\affilLL}

\date{\today}

\begin{abstract}

We demonstrate key multi-qubit quantum logic primitives in a dual-species trapped-ion system based on $^{40}$\ca and $^{88}$\sr ions, using two optical qubits with quantum-logic-control frequencies in the red to near-infrared range.  With all ionization, cooling, and control wavelengths in a wavelength band similar for the two species and centered in the visible, and with a favorable mass ratio for sympathetic cooling, this pair is a promising candidate for scalable quantum information processing.  Same-species and dual-species two-qubit gates, based on the M{\o}lmer-S{\o}rensen interaction and performed in a cryogenic surface-electrode trap, are characterized via the fidelity of generated entangled states; we achieve fidelities of 98.8(2)\% and 97.5(2)\% in \ca-\,{\ca} and \sr-\,{\sr} gates, respectively.  For a similar \ca-\,{\sr} gate, we achieve a fidelity of 94.3(3)\%, and carrying out a \sr-\,{\sr} gate performed with a {\ca} sympathetic cooling ion in a \sr-\,\ca-\,{\sr} crystal configuration, we achieve a fidelity of 95.7(3)\%.  These primitives form a set of trapped-ion capabilities for logic with sympathetic cooling and ancilla readout or state transfer for general quantum computing and communication applications.

\end{abstract}

\maketitle

\section{Introduction}

Quantum computing requires ancilla qubits as crucial components of quantum algorithmic primitives, such as quantum phase-estimation~\cite{kitaev_QPE_1995}, gate teleportation~\cite{GottesmanChuang1999}, and syndrome extraction during quantum error-correction~\cite{MikeAndIkepBook}.  In addition, particular physical implementations of quantum information processing (QIP) may utilize additional physical qubits to aid in the preparation, transport, and readout of quantum information. Particularly for trapped-ion qubits, one of the most promising quantum-computing modalities~\cite{Blatt:Wine:Nat08, MonroeKimScaling2013, BermudezAssessing2017,doi:10.1063/1.5088164}, quantum-logic operations on chains containing both computational and ancilla ions are a critical component of practical QIP systems.

In such systems, computational ions, which house the qubits primarily used for quantum logic operations, will likely be paired with ancilla ions of a different species such that control fields applied to the ancilla ions will not affect or decohere the quantum states of the computational ions~\cite{PhysRevA.61.032310,KielpinskiQCArchitecture2002}. Such ancillas could be used for sympathetic cooling after ion motion in a segmented-electrode array architecture~\cite{KielpinskiQCArchitecture2002}, for remote-entanglement generation using photons at fiber-friendly wavelengths~\cite{MonroeModularArch2014}, or for ancilla-qubit readout without decoherence of unmeasured qubits due to scattered fluorescence photons~\cite{TanMultiElement2015}. While different isotopes of the same element can provide some isolation, the isotope shifts of the relevant transitions are not typically large enough to reach levels required for high-fidelity quantum operations, much less fault tolerance.  Desired properties of ion species pairs used for QIP include a high-coherence, controllable qubit in the computational ion; similar masses of the computational and ancilla ions to allow efficient energy transfer for sympathetic cooling~\cite{PhysRevA.61.032310,WubennaSympCooling2012}; and favorable control frequencies in both species.  Control should be favorable not only in regards to the absolute frequency, but also to overlap of the frequency ranges required for the two species.  Especially in light of the potential use of integrated technologies for control-light distribution in trapped-ion quantum processors~\cite{KielpinskiIntPhotonArch2016,MehtaGrating2017,GhadimiIntegCollec2017,WestAl2O32018}, ion pairs with similar control wavelengths in the visible to near-infrared portion of the spectrum may be preferable in order to minimize coupling and propagation losses in the optical components~\cite{SoraceAgaskarSPIE2018}, while also keeping the number of different material systems needed to work across the required wavelength range to a minimum.

Here we demonstrate a set of quantum-logic primitives, using the $^{40}$\ca/$^{88}${\sr} two-species system, that form the basis for a possible QIP architecture.  Each of these ions houses a long-lived, optical-frequency qubit that has proven to be a workhorse for demanding QIP experiments and demonstrations~\cite{BenhelmMSGate2008,akerman2015universal}.  Their mass ratio near 2 allows for efficient momentum transfer for sympathetic cooling.  Furthermore, the wavelengths required for production, cooling, logic, and readout in each species fall in the optical and near infrared range, and their respective ranges have a high degree of overlap, minimizing required additional considerations for optical materials, etc., in the case of dual-species operation.  Moreover, using these two optical qubits, the frequencies of control fields required for quantum logic, where the highest intensities and phase stability are required, are at very favorable wavelengths (729~nm and 674~nm for {\ca} and \sr, respectively) for laser and optical component technology.  As an additional benefit, light emitted by broad transitions in these species, as would be used for remote-entanglement generation, is in the blue and infrared parts of the visible spectrum, considerably simplifying collection, distribution, and interference of this light when compared to species with similar transitions in the ultra-violet.  Our demonstration of a suite of two-qubit quantum-logic primitives using two optical-frequency qubits helps establish this ion-species pair as a useful system for scalable-QIP explorations.

Prior work in dual-species trapped-ion quantum logic includes demonstrations of so-called quantum-logic spectroscopy~\cite{SchmidtQuantumLogicSpectroscopy} to enable the operation of optical clocks using ions with inaccessible or inconvenient transitions; these typically focus on the Al$^{+}$ clock ion with another species similar in mass used to manipulate and read out its state~\cite{RosenbandClockCompare2008}.  Entangling quantum gates have also been performed with the ion systems Be$^{+}$/Mg$^{+}$~\cite{TanMultiElement2015}, Be$^{+}$/Ca$^{+}$~\cite{negnevitsky2018repeated}, $^{40}$Ca$^{+}$/$^{43}\mathrm{Ca}^{+}$~\cite{BallanceHybridLogic2015}, and Ba$^{+}$/Yb$^{+}$~\cite{PhysRevLett.118.250502}.  While these pairs will likely find particular application, they all have drawbacks, such as technologically challenging wavelengths, insufficient separation in internal-state energy splittings, or large masses, and so it is beneficial to explore alternative dual-species systems that may have complementary strengths.  

The \ca/{\sr} system has been considered recently for applications in QIP, though few quantum logic operations in the combined system have been reported.  For instance, these species form the basis of several analyses of large-scale quantum computing platforms, both as sympathetic cooling ancillas~\cite{BermudezAssessing2017} and for photon-emitting intermediaries in optically linked architectures~\cite{Nigmatullin_2016}; a method has also been suggested to perform an interspecies gate in the \ca/{\sr} system using a single laser wavelength~\cite{BallanceHybridLogic2015}.  While considerable work demonstrating same-species, two-qubit operations in \ca~\cite{BenhelmMSGate2008,KirchmairHotGates2009,PhysRevLett.117.060504_2016,HartyNearFieldMicrowaves2016,schafer2018fast}, and to a lesser extent \sr~\cite{akerman2015universal,shapira2018robust}, exists, their use together has not been investigated widely.  Due to its potential utility, it is important to explore the implications of quantum operations based on this species pair.

\section{Architectural Components}
\label{sec:arch}

\begin{figure}[t !]
\includegraphics[width = 1.0 \columnwidth]{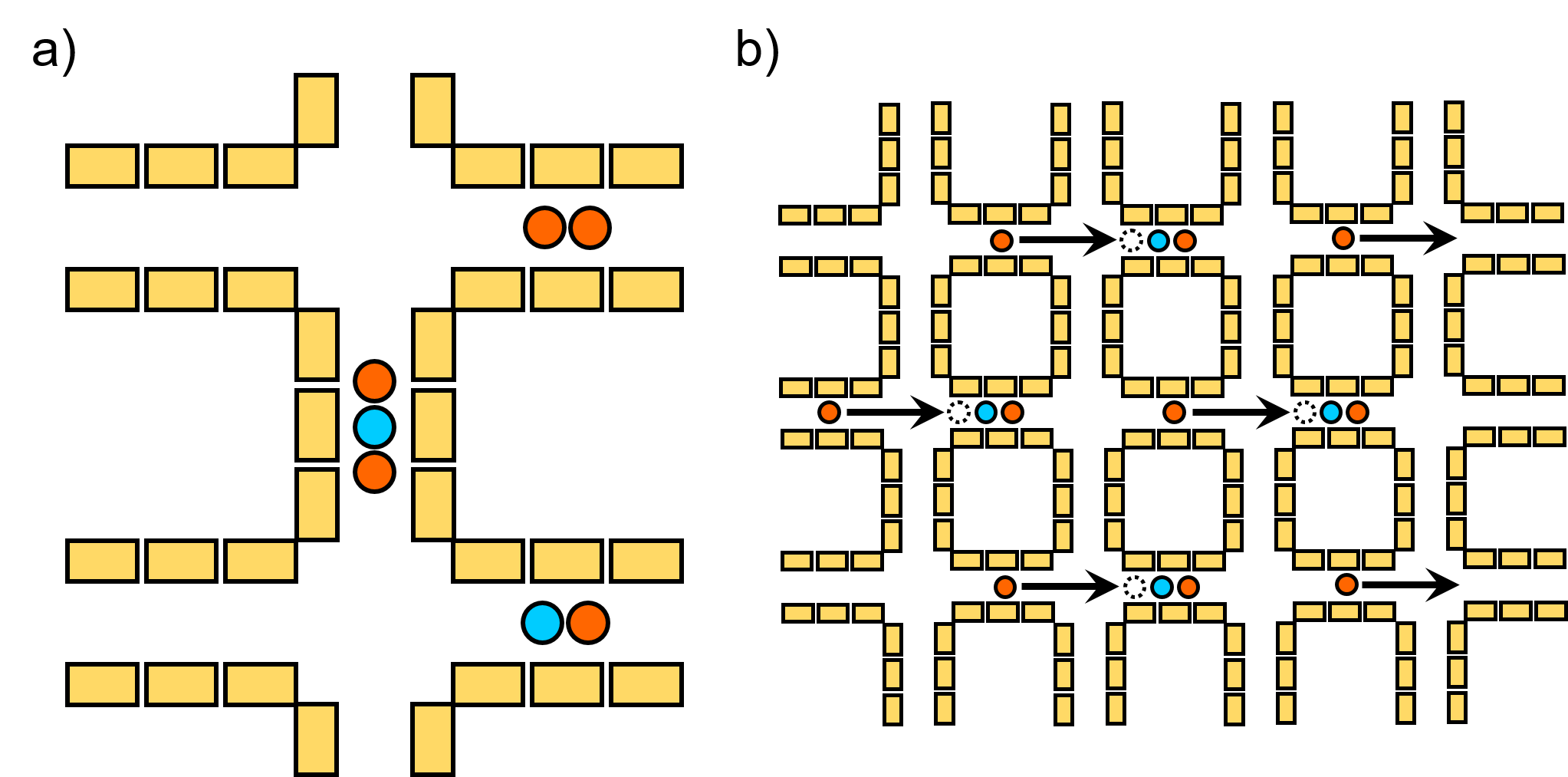}
\caption{Trap array architecture with ion-transport-based connectivity. (a) Subsection of array showing segmented electrodes which create both static potential wells at multiple array sites and dynamically variable potentials for ion transport.  Different zones depict various crystal configurations of computational ions (red) and ancilla ions (blue). (b) A larger section of such an array configured for quantum computation with surface-code error correction encoding; in this case two-qubit gates are perfomed between nearest-neighbor computational ions after transporting lone computational ions to the zones (alternating in a checkerboard pattern) where a computational ion is housed with an ancilla.  One step in the error-correction cycle is depicted.  The ancilla is used to prepare the shared motional state before gate operations, and the ancilla could also be used for periodic syndrome readout without detrimental effect on unmeasured computational ions due to photon scattering.}
\label{fig:arraytraps}
\end{figure}

One potential architecture for QIP with trapped ions consists of a two-dimensional array of trapping zones, interconnected via transport regions, in which two species of ions are held~\cite{theBible,KielpinskiQCArchitecture2002} (see Fig.~\ref{fig:arraytraps}a).  Each trapping zone holds a few-ion, one-dimensional ion crystal, in which multiqubit operations can be performed.  Ions are moved between trapping zones via the transport regions to bring ions from different array sites into the same crystal, such that a high degree of connectivity of multi-qubit operations can be maintained across the array, limited only by the complexity of the transport region network.  Housing 1--4 ions in each site simplifies the vibrational mode spectrum when compared to keeping all ions in one crystal, while potentially permitting simultaneous individual addressing of many ions throughout an array.  Ion crystals composed of ions from separate array sites, transported and then joined together, may acquire vibrational-mode excitation that can limit gate fidelity.  Hence, any array site where multiqubit gate operations will be performed contains one or two ancilla ions used primarily for sympathetic cooling prior to gate operations, allowing preparation of select vibrational modes of the crystal without affecting the internal state of the computational ions.

Such an architecture could be flexible in terms of application.  The segmented trap structure can be tailored in terms of connectivity from nearest-neighbor, e.g. for surface-code quantum-error correction or quantum emulation of solid-state Hamiltonians; to fully connected, e.g. for quantum chemistry simulations requiring Jordan-Wigner transformations between qubit and orbital bases.  In all these cases, however, the composition of each array site can be essentially identical, able to maintain one to two of each of the computational ions and ancilla ions in a linear crystal.  We focus here on demonstrations of key two-qubit primitives that enable operations within these individual sites, e.g. architectural components useful for multiple applications, using the \ca/{\sr} system.

Basic two-qubit logic between ions of the same species forms the foundation of any QIP application that requires entanglement or qubit interaction, and we therefore begin with \ca-{\ca} and \sr-{\sr} two-qubit gates which set the baseline capability of this system.  These gates lead naturally to an architectural primitive in which a gate is performed between two ions of a single species in the presence of a sympathetic coolant ancilla of a second species.  This operation would be the primary multi-qubit gate in a case in which all array sites contain a single computational ion, while a subset of the sites also contain a co-located coolant ancilla.  To perform logic operations between pairs of computational ions, a lone computational ion is transported from its home site to the home site of a computational ion which houses an ancilla.  All three ions are then joined in a single crystal.  The ancilla is used to remove any unwanted motional excitation accrued during ion movement and crystal merging~\footnote{The opposite operation, crystal separation, will in general occur before the next gate involving one of the current computational ions, and sympathetic cooling via an ancilla subsequent to transport and merging will also remove the kinetic energy acquired during this process.}; the gate is then performed between the computational ions.  As a particular example, the surface code~\cite{FowlerSurfaceCode2012} can be implemented in such an architectural scheme in which half the sites, in a checkerboard pattern, of a square array contain sympathetic-cooling ancillas (see Fig.~\ref{fig:arraytraps}b).  Here we demonstrate each of the above-mentioned same-species, multi-ion, quantum-logic architectural components.

\begin{figure}[t !]
\includegraphics[width = 0.95 \columnwidth]{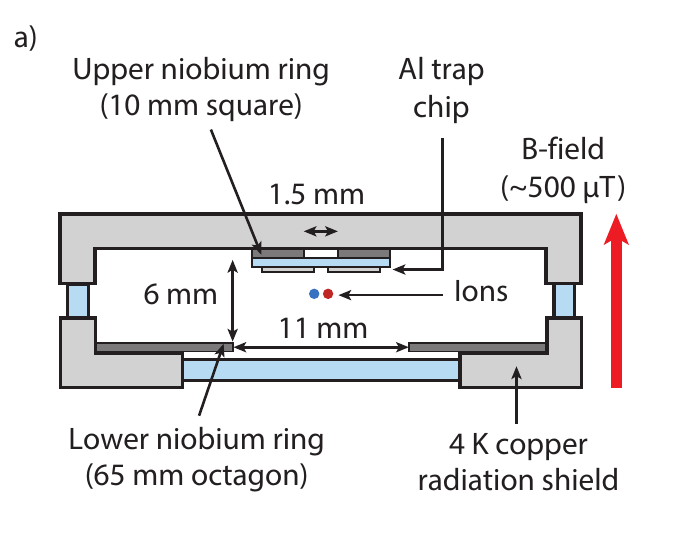}\\
\includegraphics[width = 1.0\columnwidth]{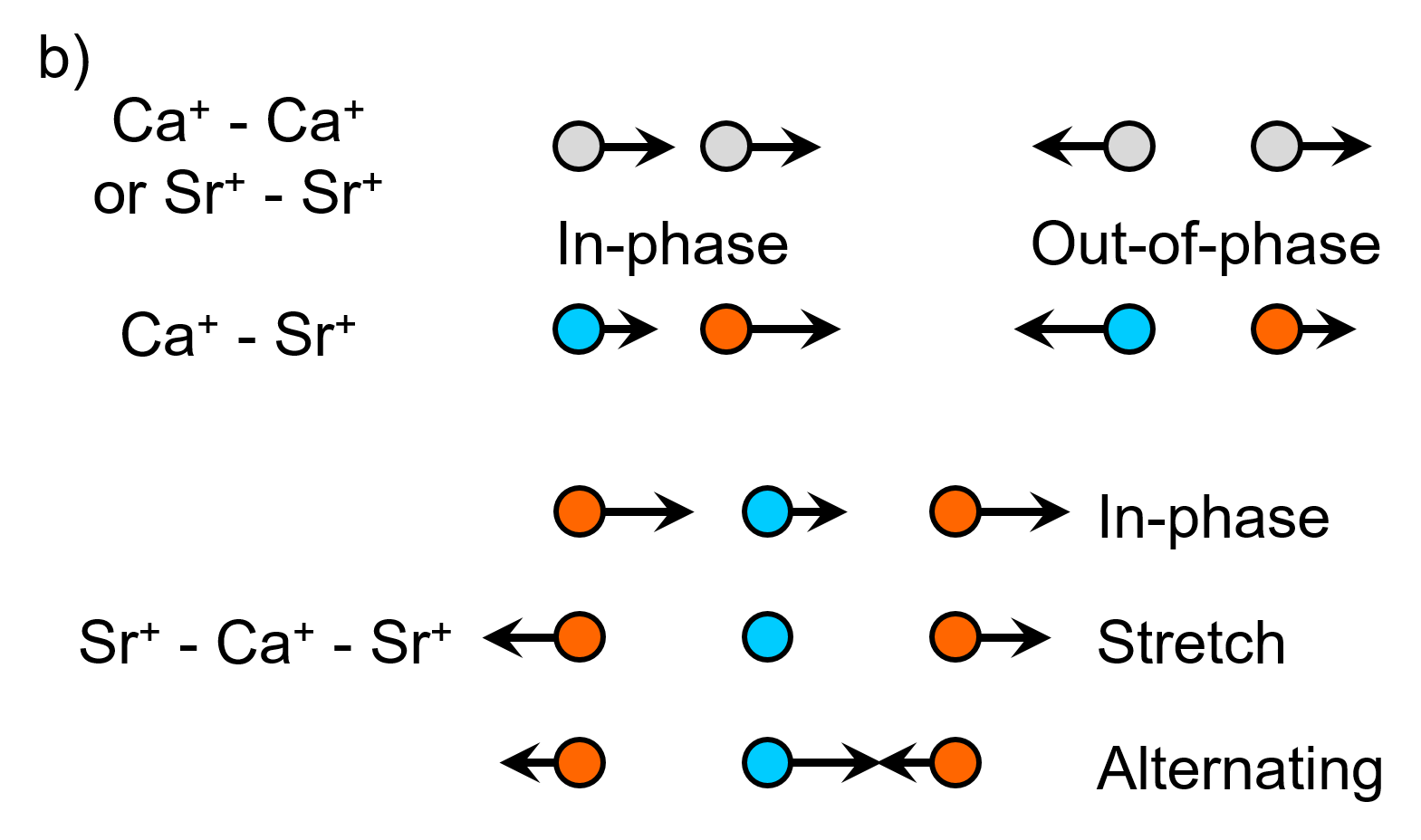}
\caption{Magnetic-field shielding and ion-crystal motional modes. (a) Self-shielding, superconducting ring geometry for suppressing magnetic-field noise. Fluctuations in the magnetic flux threading the rings induce persistent supercurrents that (partially) cancel the changes in magnetic flux. (b) Ion-crystal configurations and axial normal vibrational modes pertaining to two- and three-ion chains employed in this work. Arrows (not drawn to scale) indicate the relative amplitudes and directions of the normalized motional eigenvectors for the different ion crystals.}
\label{fig:nbrings}
\end{figure}

Dual-species operations provide additional key capabilities for trapped-ion-based quantum information processors, and therefore form other primitives of interest.  For instance, in a larger-scale system employing quantum error correction, subsets of the qubits must be measured during the computation.  Moreover, measurement-based quantum computing also requires projecting a subset of qubits while maintaining coherence of the unmeasured remainder.  In both cases, inter-species transfer of state population to ancillas just prior to measurement can avoid decoherence in computational ions due to resonant light scattered from measured ions in close proximity~\cite{BruzewiczQLAR2017}.  Another important example is repeat-until-success remote-entanglement generation for linking modules of a composite quantum information processor~\cite{PhysRevLett.118.250502}; in this case, ions desired for long-distance communication, either due to particular wavelengths or beneficial level structures, may be different from the computational species.  Transfer of the quantum state from a computational ion to one qubit of a Bell-entangled ancilla pair can allow for independent choices of ion species in these roles.

Transfer to ancillas to avoid decoherence during measurement does not necessarily need to preserve qubit phase information, and hence techniques based on quantum-logic spectroscopy~\cite{SchmidtQuantumLogicSpectroscopy,HumeAdaptiveDetection2007} can be utilized to accomplish this readout scheme.  A full phase-preserving state swap, e.g a series of three CNOT gates based on a dual-species entangling gate, can also fulfill this task~\cite{TanMultiElement2015}; for state transfer into part of a remote entangled pair as a prerequisite for additional computation, only full quantum state transfer will suffice.  We have previously performed quantum-logic-assisted readout of a {\ca} ion using a {\sr} ion~\cite{BruzewiczQLAR2017}, a technique that may be useful for resonant-light-scatter-free syndrome extraction in a close-packed array.  This operation required a pair of $\pi$-pulses on the sideband transitions corresponding to a shared vibrational mode after it was cooled to the ground state.  Here we demonstrate a more general dual-species MS entangling gate between {\ca} and \sr, a method that has an advantage beyond phase preservation in that ground-state cooling is generally not required for MS gates~\cite{KirchmairHotGates2009}.  It could therefore be used for both syndrome extraction and remote-entanglement-generation applications.

\section{Experimental Methods}

We confine and manipulate $^{40}${\ca} and $^{88}${\sr} ions in a linear surface-electrode trap consisting of a 2~$\mu$m-thick, patterned aluminum layer on a sapphire substrate, similar to traps used in previous work \cite{PhysRevA.89.012318_2014,PhysRevA.98.063430}.  Single-ion axial trap frequencies in the $0.5$--$2$~MHz range are produced via application of potentials to segmented electrodes defined along the axial direction; radial trapping at frequencies near 5~MHz is produced through application of a radio-frequency (RF) potential near 50~MHz to a subset of the electrodes.  Ions are held 50~$\mu$m from the trap-chip surface in a cryogenic vacuum chamber achieving ultra-high vacuum pressure in which the trap is maintained at a temperature below 6~K.  The ions are loaded into the trap via photoionization from neutral atomic beams produced via acceleration from co-located Ca and Sr two-dimensional magneto-optical traps to the remotely located trap chip~\cite{Bruzewicz2016,BruzewiczQLAR2017}.

Environmental magnetic-field fluctuations are suppressed using a pair of superconducting niobium rings~\cite{PhysRevA.81.062332} located above and below the trap chip. The rings are centered radially on the ion location, with one 10-mm-square ring attached just beneath the trap's sapphire substrate and one 65-mm-diameter octagonal ring attached to the radiation shield approximately 5~mm from the trap surface; this design is shown schematically in Fig.~\ref{fig:nbrings}.  Induced supercurrents in the rings compensate variation in the local magnetic field~\cite{gabrielse1991superconducting} in the direction of an applied, axial quantizing field of approximately $5\times10^{-4}$~T.  In practice, the quantizing magnetic field is produced using the supercurrent induced in the rings.  To reduce field noise, the current in coils used to inject this supercurrent is removed once the rings are in the superconducting state.  We have measured suppression of slow magnetic-field fluctuations by a factor of 17--20~dB using this technique.

Qubits are defined using the electronic states in each ion, with an optical-frequency separation between the $|0\rangle\equiv |(n-1)\ {}^{2}D_{5/2}, m_{J}=-5/2\rangle$ and $|1\rangle\equiv |n\ {}^{2}S_{1/2}, m_{J}=-1/2\rangle$ states, where $n=\{4,5\}$ for \{\ca,\ \sr\}.  Light for qubit manipulation and multi-qubit operations is derived from separate systems each consisting of an external-cavity-stabilized diode laser, frequency locked to an ultra-low-expansion (ULE) glass cavity.  Transmitted light, filtered by the cavity, is used to injection-lock one or more slave diode lasers~\cite{akerman2015universal}, and the output is amplified via one or more tapered optical amplifiers.  Light is passed through several acousto-optic modulators (AOMs) to shift the frequency and modulate the amplitude and phase (and frequency for multi-qubit operations) before being delivered to the ions through windows in the vacuum chamber.

Experimental trials each begin with Doppler cooling of the ions using light at 397~nm (422~nm) in conjunction with repumping light at 866~nm (1092~nm) for \ca (\sr; the same construction will be used throughout this paragraph).  Light at 854~nm (1033~nm) is also applied to quench any residual population in $|0\rangle$ through the $n\ {}^{2}P_{3/2}$ level.  Ions are Doppler cooled for approximately 1~ms, after which resolved-sideband cooling is performed to bring the ions to the motional ground state for a subset of the axial vibrational modes of the ions in the crystal.  For same-species, two-qubit gates with two ions in the crystal, sideband cooling pulses at 729~nm (674~nm) interspersed with quenching pulses at 854~nm (1033~nm) are used to bring the in-phase (IP) and out-of-phase (OOP) modes (see Fig.~\ref{fig:nbrings}b) to average occupation below approximately 0.05.  Sideband cooling for dual-species, two-qubit gates and single-species, three-ion, two-qubit gates will be described below in the respective sections.  Optical pumping after sideband cooling serves as state preparation, bringing the ions to $|1\rangle$ using a combination of 729~nm and 854~nm (674~nm and 1033~nm) light.  After state-preparation, bichromatically modulated light is used to perform two-qubit entangling operations via the M{\o}lmer-S{\o}rensen (MS) technique~\cite{MolmerSorensenGate}.

State detection is performed by applying the wavelengths used for Doppler cooling, but without the quench light, such that an ion in $|1\rangle$ will fluoresce at 397~nm (422~nm) while an ion in $|0\rangle$ will not.  Fluorescence is collected using a high-numerical-aperture objective outside the vacuum chamber and directed to an electron multiplying CCD or photo-multiplier tube (PMT) for imaging or state detection, respectively.  For single-species gates, ions are detected simultaneously, and the detection time, typically a few milliseconds, is set such that experimental photon-number histograms corresponding to 0, 1, or 2 ions in the scattering $|1\rangle$ state are sufficiently separated to allow discrimination between these cases with error probability below approximately 0.001.  Note that of the four possible two-qubit state outcomes, two of them are indistinguishable when measured using the non-imaging PMT; hence only the sum of the probabilities for the states $|01\rangle$ and $|10\rangle$ are measured.  This is not a limitation for determination of the fidelities of the created Bell states, since only the populations of the $|00\rangle$ and $|11\rangle$ states, and the parity of the two-qubit state measured in an auxiliary experiment performed on the Bell state, are needed~\cite{Sackett4IonEntanglement2000}. In the mixed-species gate, fluorescence is detected from each ion sequentially, but the fidelity is calculated in the same way as in the single-species gates.

\section{M{\o}lmer-S{\o}rensen Logic Gates and Considerations Pertaining to Ion Crystal Configuration and Motional Modes}


The MS gates demonstrated here are enacted through optical dipole forces applied at frequencies near a particular shared motional mode of the trapped-ion crystal.  When bringing about these forces via a bichromatic light field with frequency components detuned by $\delta$ above and below the blue and red motional sidebands (mode frequency $\omega_{\beta}$), the interaction Hamiltonian for two ions is~\cite{Roos_2008_amp_mod_gates}

\begin{widetext}
\begin{equation}
    H_{I}(t)=\hbar \Omega \left(e^{-i (\omega_{\beta}+\delta) t}+e^{i (\omega_{\beta}+\delta) t}\right)  e^{ i\eta ( a e^{-i \omega_{\beta} t} + a^{\dagger} e^{i \omega_{\beta} t}) } \left(\sigma_{+}^{(1)} + \sigma_{+}^{(2)}\right) + \textrm{h.c.}
\end{equation}
\end{widetext}

\noindent Here $a$ is the annihilation operator of the vibrational mode of interest, $\Omega$ is the $|1\rangle \rightarrow |0\rangle$ transition Rabi frequency, $\sigma_{+}^{(j)}$ is the raising operator for the electronic spin qubit of ion $j$ defined via the Pauli spin operators as $\sigma_{+}=(\sigma_{x}+i \sigma_{y})/2$, and $\eta=k\, z_{\textrm{RMS}}$ is the Lamb-Dicke parameter which expresses the ratio between the vibrational ground-state wavefunction size $z_{\textrm{RMS}}=\sqrt{\hbar/(2 m \omega_{\beta})}$ and the spatial gradient of the electromagnetic field, as expressed by the mode-direction projection of the wavevector $k$ of the light used to drive the transition.  We are here assuming a single-species MS gate for simplicity; below we generalize the Lamb-Dicke parameter to more complex configurations.   Taking the rotating-wave approximation and in the Lamb-Dicke limit ($\eta \sqrt{\langle (a+a^{\dagger})^{2} \rangle} \ll 1 $), this Hamiltonian becomes

\begin{equation}
    H_{I}(t)=-\hbar \eta \Omega ( a e^{-i \delta t} + a^{\dagger} e^{i \delta t})  \left(\sigma_{y}^{(1)} + \sigma_{y}^{(2)}\right).
\end{equation}

\noindent  This interaction couples the motional mode to the joint spin state of the two ions, and by driving off-resonance with detuning $\delta$, the vibrational mode acquires a geometric phase that depends on the joint spin state as it traverses a curved path in the phase space of the mode.  At drive times equal to multiples of  $t_{g}=2\pi/\delta$ the curved path becomes a closed loop and the ions' spin and motion are disentangled, while a subset of the two-ion spin states acquire a phase relative to the others; by setting the interaction strength such that $\eta\Omega=\delta/4$, a maximally (spin) entangled state can be created after a time $t_{g}$.

The mode structure affects the gate time through the Lamb-Dicke parameter, but this is rather straightforward in same-species gates.  We now discuss considerations affecting the gate time for dual-species operations, where the Lamb-Dicke parameter is ion-dependent, and we consider errors due to excitation of spectator modes spectrally close to the gate-drive frequency, an effect that can be exacerbated in mixed-species systems.  See the Appendix for the calculation of the mode structure in dual-species crystals.

In general, it is desirable to execute multi-qubit operations in the shortest possible time with the highest possible fidelity, and both of these design goals can be affected by the chosen motional mode used to execute the gate. For the optical qubits used here, the attainable sideband Rabi frequencies, which set the MS gate time, are determined by the  electric quadrupole transition matrix elements between the chosen qubit states, the Lamb-Dicke parameters for the relevant motional mode, and the qubit laser intensities. These parameters can easily be made, to a very good approximation, equal for ions of the same species. In multi-species crystals, however, the values can vary significantly, and each must be well characterized in order to achieve high-fidelity operation. For example, the electric quadrupole transition matrix element for the chosen qubit states in $\mathrm{Ca}^{+}$ is approximately 0.7 times that of $\mathrm{Sr}^{+}$~\cite{PhysRevA.95.042507}.  As shown in the Appendix, the Lamb-Dicke parameters vary for different ion species in different motional modes using different crystal configurations. The sideband Rabi frequencies can be equalized, however, by adjusting the gate-laser intensities at the ion locations, subject to the constraints imposed by the available laser power and the achievable beam waists.  We also point out that the inclusion of a second atomic species in three-ion crystals brings about additional considerations, as each ion crystal configuration yields different normal mode frequencies and motional amplitudes.


In addition to affecting the gate speed, motional mode choice can also determine the sensitivity of the ion chain to different types of electric-field noise. Homogeneous electric fields couple most strongly to in-phase motional modes. Hence, some level of motional state heating suppression can be expected for modes with out-of-phase ion motion~\cite{Wineland1998,PhysRevA.61.032310}. A single {\ca} ion heating rate of $8.6\pm0.4$~quanta/s for a $2\pi\times1.94$~MHz trap frequency was measured in the trap used here, and while the heating rates of in-phase axial modes for multi-ion chains are expected to be slightly larger than this, they are not currently limiting gate fidelity.

Lastly, the excitation spectrum becomes increasingly dense in larger ion crystals, especially when including the effects of higher order motional sidebands. Nearby transitions can lead to unwanted state couplings and subsequent gate errors. This consideration is especially pertinent in the $^{40}\mathrm{Ca}^{+}$/$^{88}\mathrm{Sr}^{+}$ system where the mass ratio $\mu=2.2$ gives rise to a number of near degeneracies, as shown in the Appendix. 

A key example is the small energy splitting between the axial out-of-phase mode first-order sideband and the axial in-phase mode second-order sideband in this pair, due to the mode frequency ratio $\omega_{\textrm{OOP}}/\omega_{\textrm{IP}}=1.988$.  Driving the gate on the OOP mode is desirable in some cases since it is affected less than the IP mode by motional heating.  For a typical axial trap frequency of $\omega_{\textrm{IP}}\approx 2\pi \times 1$~MHz, the OOP-2IP splitting is only $2\pi\times 12$~kHz, approximately equal to typical detunings $\delta$ from the motional sidebands used during gate operations.  Thus a detuned drive of the OOP mode may be near resonant with the second-order IP sideband, potentially leading to error from unwanted displacement of this mode's motional state.  This error is similar to off-resonant excitation of spectator modes on the first-order sideband~\cite{BallanceThesis2014}, though it depends on the different carrier Rabi frequencies and mode-dependent Lamb-Dicke parameters for each species and has an extra factor of the Lamb-Dicke parameter due to the higher-order excitation.  We therefore expect the error to be (for small displacements and mode heating rates, and ignoring other off-resonant terms)

\begin{equation}
\epsilon_{2\times \textrm{IP}} \approx |\alpha|^{2}\, \left(\bar{n}_{\textrm{IP}}+\frac{1}{2}\right)
\end{equation}

\noindent where $\bar{n}_{\textrm{IP}}$ is the average occupation of the IP mode before the gate and $\alpha$ is the displacement of the OOP mode due to the drive, expressed as (assuming in-phase excitation of the two ions at frequency $\omega$)

\begin{equation}
\alpha =\frac{1}{2} \int_{0}^{t_{g}} \left( \eta_{\textrm{Ca,IP}}^{2} \Omega_{\textrm{Ca}} + \eta_{\textrm{Sr,IP}}^{2} \Omega_{\textrm{Sr}} \right) e^{-i\, (\omega-2\omega_{\textrm{IP}})\, t}\ dt.   
\end{equation}

\noindent Here $\eta_{j,\beta}$ and $\Omega_{j}$ are the Lamb-Dicke parameter (see the Appendix for the definition in the multi-species case) and carrier Rabi frequency, respectively, for ion $j$ in mode $\beta$, and $t_{g}$ is the MS gate time.  Evaluating this integral, we obtain

\begin{equation}
|\alpha|^{2} = \frac{\left( \eta_{\textrm{Ca,IP}}^{2} \Omega_{\textrm{Ca}} + \eta_{\textrm{Sr,IP}}^{2} \Omega_{\textrm{Sr}} \right) \sin^{2} [ (\omega-2\omega_{\textrm{IP}})\frac{t_{g}}{2}]}{(\omega-2\omega_{\textrm{IP}})^{2}}.
\label{eq:disp}
\end{equation}

\noindent For the worst case scenario, on-resonant driving of the IP second order sideband ($\omega=\omega_{\textrm{OOP}}+\delta= 2\omega_{\textrm{IP}}$ for the drive near the blue sideband), and for the dual-species gate conditions $\eta_{\textrm{Ca,OOP}} \Omega_{\textrm{Ca}}=\eta_{\textrm{Sr,OOP}} \Omega_{\textrm{Sr}}=\delta / 4$, $\alpha= (\delta t_{g} / 8)(\eta_{\textrm{Ca,IP}}^{2} / \eta_{\textrm{Ca,OOP}} + \eta_{\textrm{Sr,IP}}^{2} / \eta_{\textrm{Sr,OOP}})$; the error for a gate which acquires a $\pi$ phase due to displacement around one-loop on the OOP mode ($t_{g}=2\pi / \delta$) would then be

\begin{equation}
\epsilon_{2\times \textrm{IP}} =  \frac{\pi^2}{16} \left(\bar{n}_{\textrm{IP}}+\frac{1}{2} \right) \left( \frac{\eta_{\textrm{Ca,IP}}^{2}}{\eta_{\textrm{Ca,OOP}}} + \frac{\eta_{\textrm{Sr,IP}}^{2}}{\eta_{\textrm{Sr,OOP}}} \right)^{2}.
\end{equation}

\noindent  Even assuming negligible initial population in the motional modes, this error can be substantial; for a $\omega_{\textrm{IP}}= 2\pi \times 1$~MHz axial mode in the ground state, the error would be $\epsilon_{2\times \textrm{IP}} = 0.013$ for a $^{40}\mathrm{Ca}^{+}$--\ $^{88}\mathrm{Sr}^{+}$ crystal.  Moreover, as can be seen from the functional form of the acquired displacement (Eq.~\ref{eq:disp} with $\omega=\omega_{\textrm{OOP}}+\delta$), the width in gate detuning of the effect of driving near this IP resonance is set by $2\omega_{\textrm{IP}}-\omega_{\textrm{OOP}}$, so small relative changes in detuning will not be effective in significantly reducing this error.  It is therefore prudent to avoid such near coincidences, either via judicious choices of detuning and laser intensity if available (though this removes a degree of freedom often used to optimize gate operation), trap frequency, mode of operation, or isotopes of the ions in the chain (see the Appendix for the mode frequency ratios for several different isotope combinations of {\ca} and {\sr}); the latter can be very effective since the absolute mode frequency difference sets the relevant scale.  In the work presented here, the modes used to execute the dual-species MS gates were chosen to be spectrally well separated from all other transitions for attainable gate speeds.

\begin{figure*}[t !]
\parbox[b][1.93 in][t]{0.1 in}{a)}\includegraphics[width = 0.65 \columnwidth]{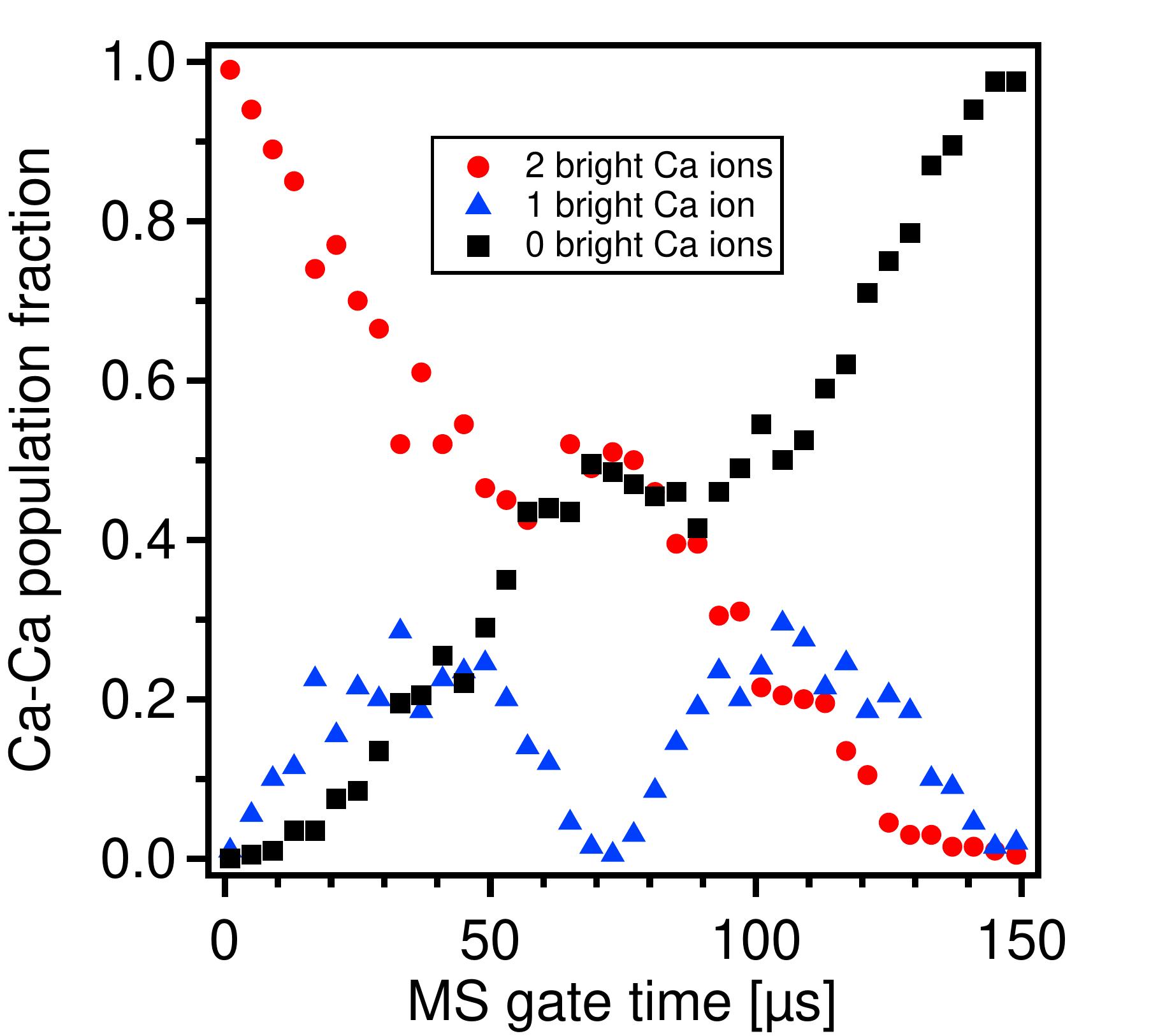}
\parbox[b][1.93 in][t]{0.1 in}{b)}\includegraphics[width = 0.65 \columnwidth]{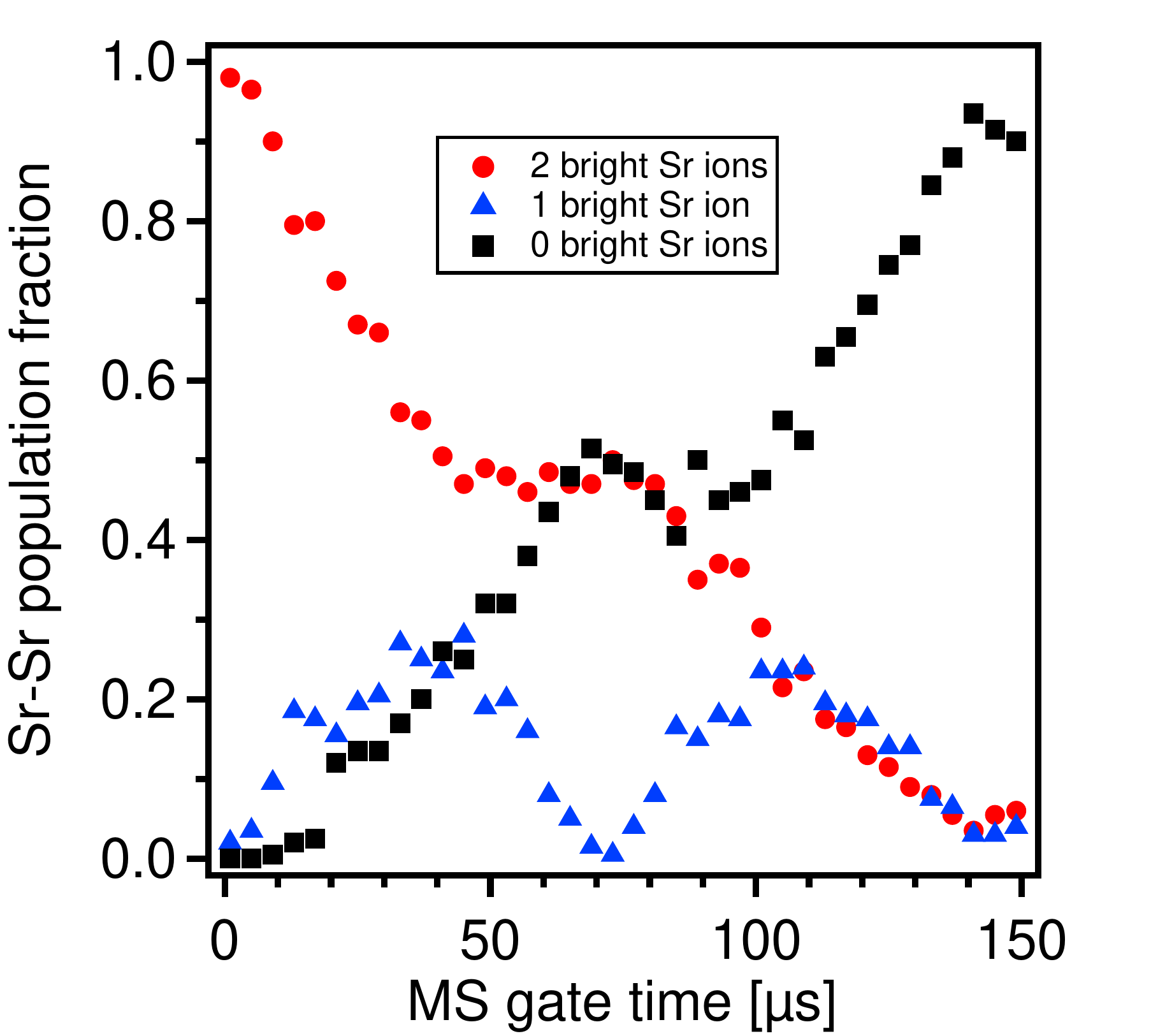}
\parbox[b][1.93 in][t]{0.1 in}{c)}\includegraphics[width = 0.65 \columnwidth]{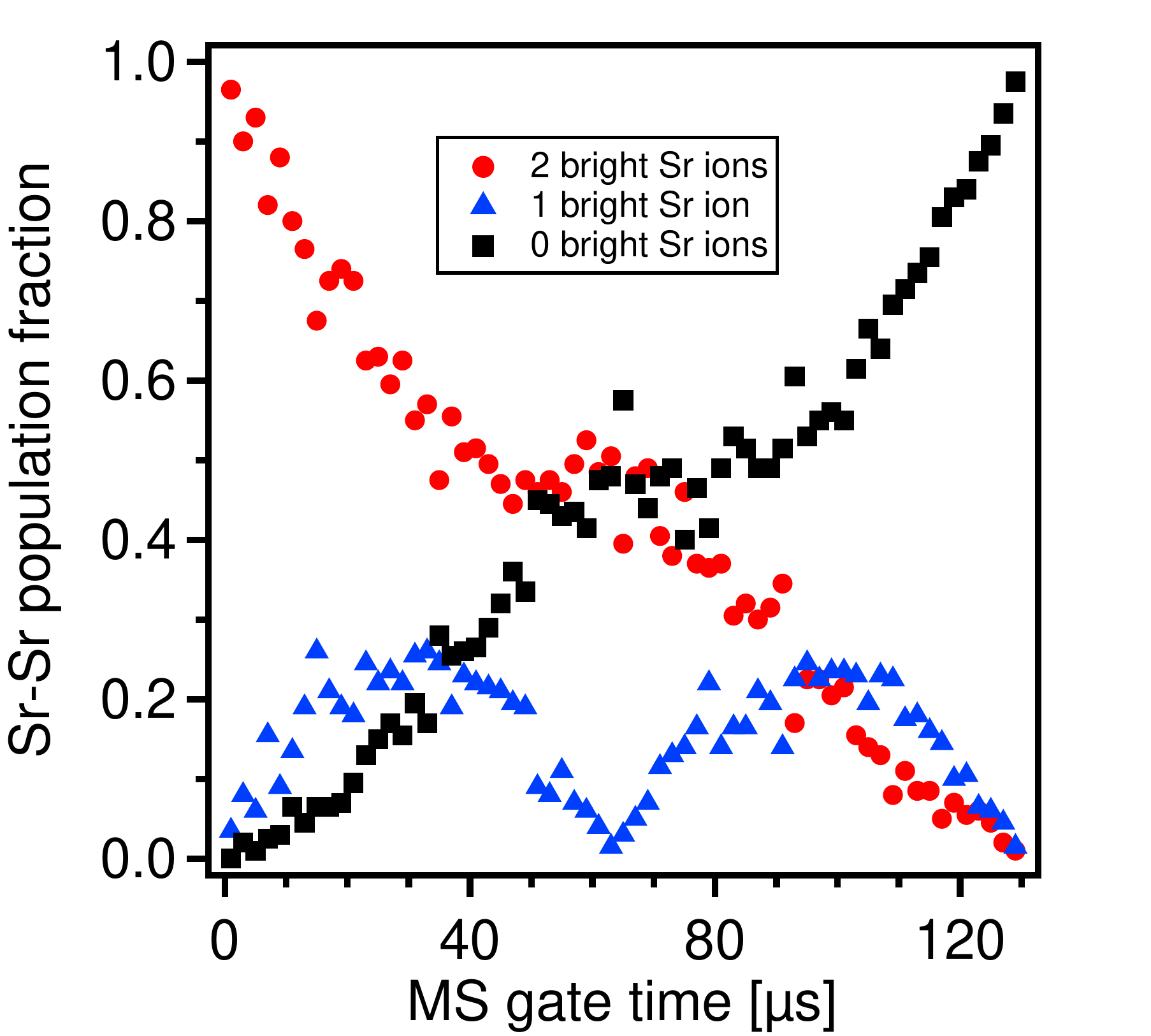}\\
\parbox[b][1.58 in][t]{0.1 in}{d)}\includegraphics[width = 0.65 \columnwidth]{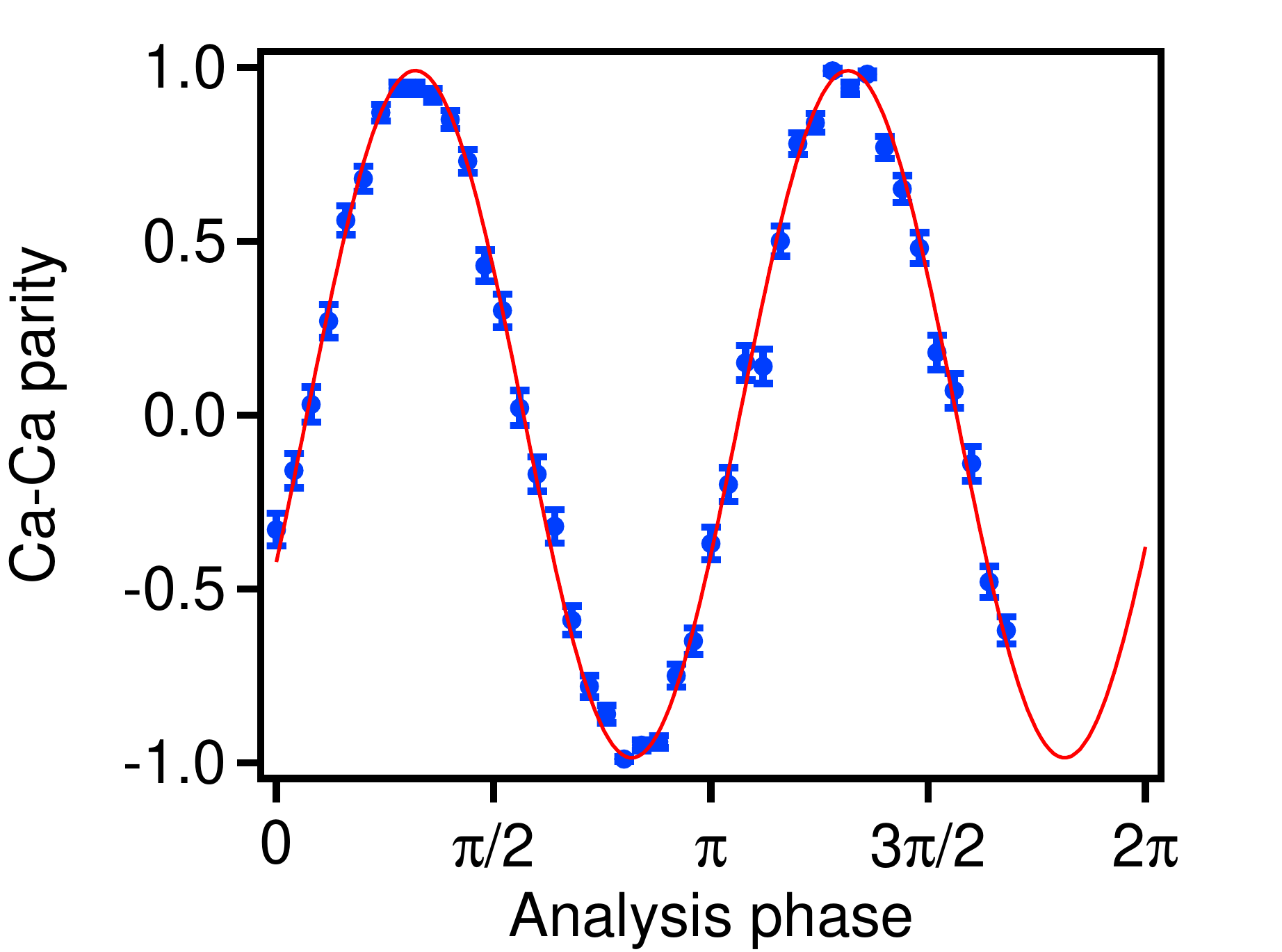}
\parbox[b][1.58 in][t]{0.1 in}{e)}\includegraphics[width = 0.65 \columnwidth]{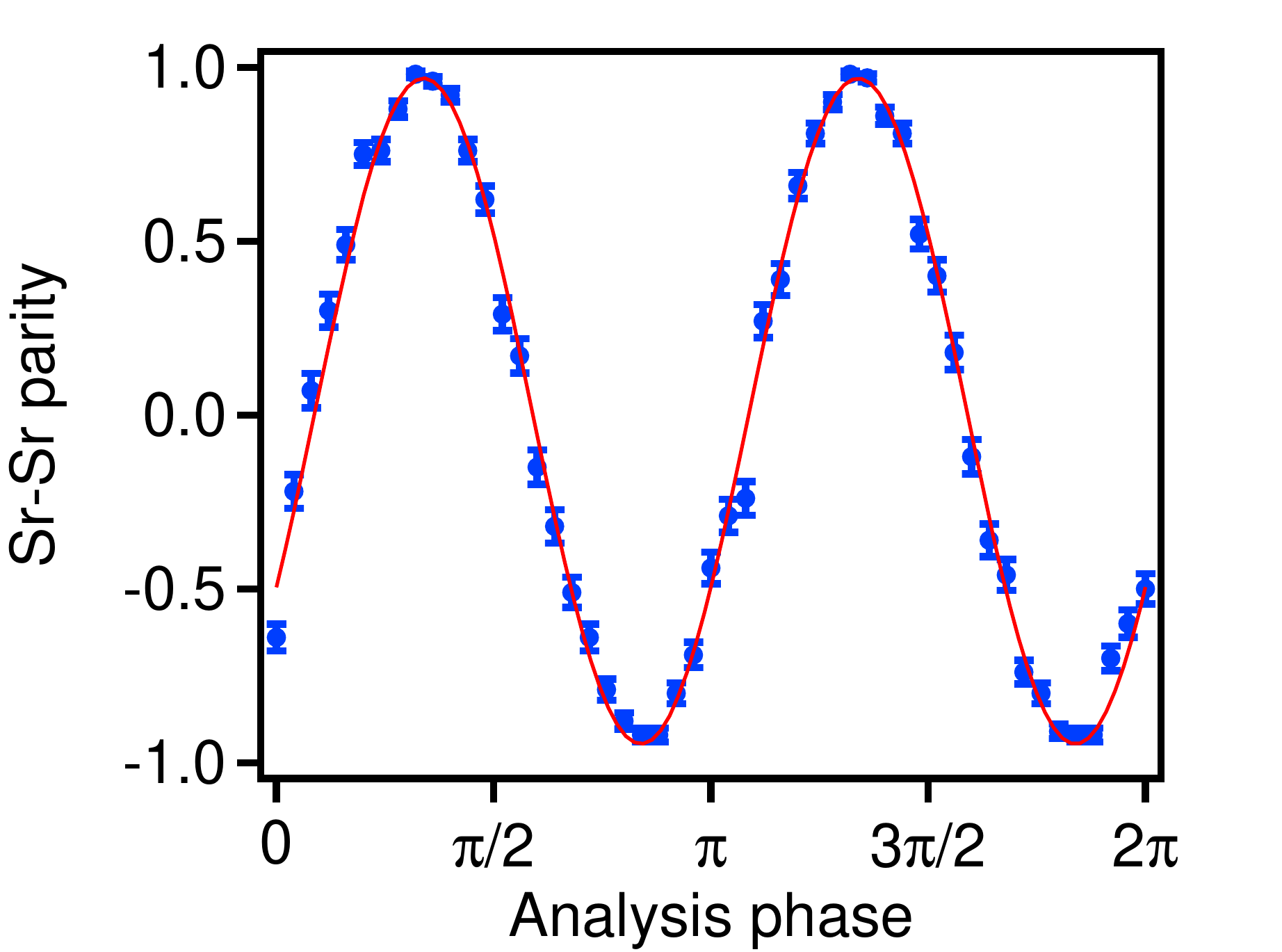}
\parbox[b][1.58 in][t]{0.1 in}{f)}\includegraphics[width = 0.65 \columnwidth]{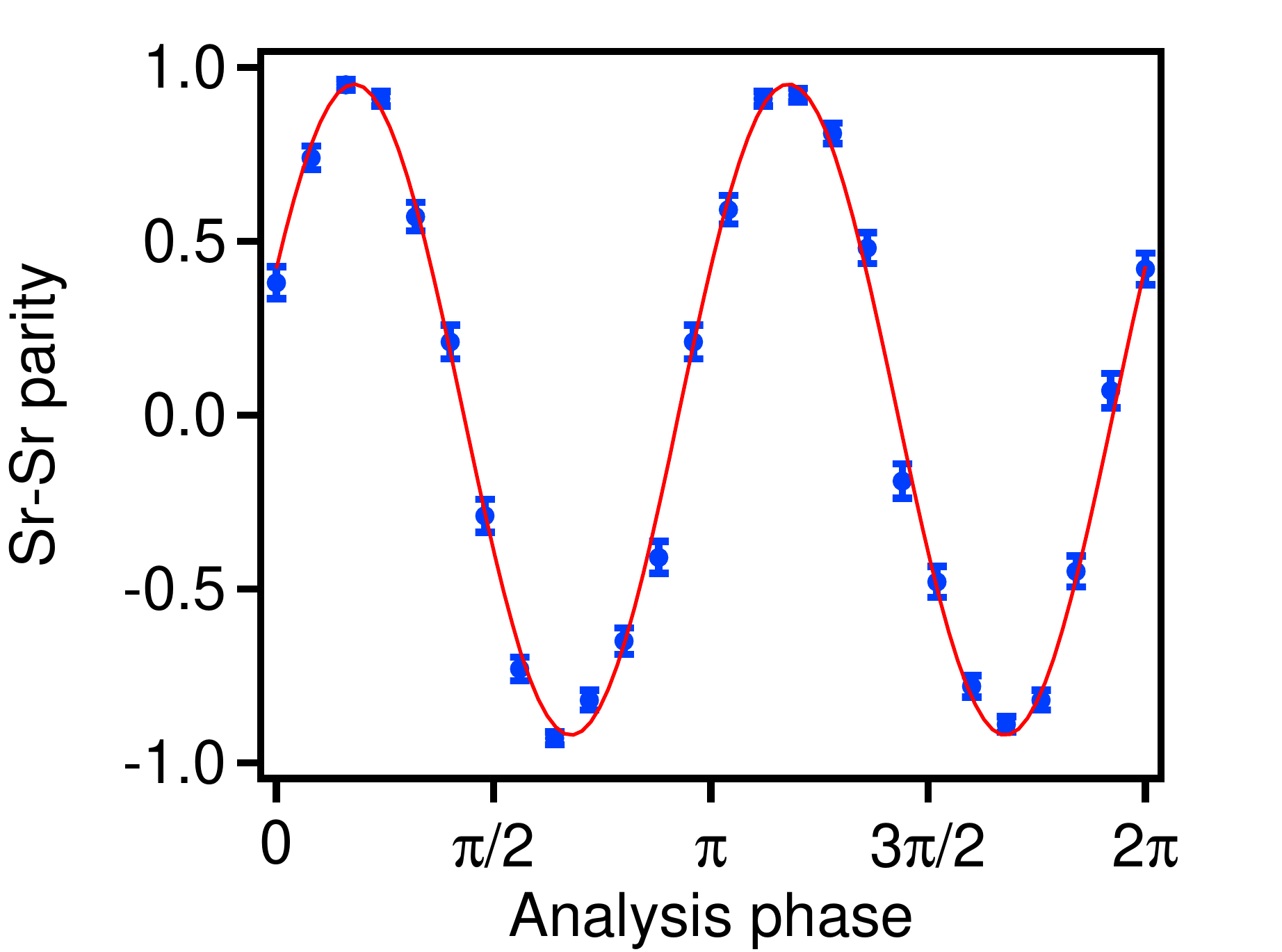}
\caption{Two-qubit entangling gates with ions of the same species.  (a) \ca--\ca M{\o}lmer-S{\o}rensen (MS) gate performed on the in-phase (IP) mode at $2\pi\times1.2$~MHz; here the measured state populations are plotted as a function of duration of the gate pulse.  Starting with the product state $|11\rangle$ (two bright ions) at time 0, a maximally entangled state is created after 71~$\mu$s. (b) \sr--\sr MS gate performed on the IP mode at $2\pi\times1.3$~MHz; the Bell state is created at a duration of 72~$\mu$s.  (c) MS gate on two \sr ions with initial sympathetic cooling performed using a central \ca ancilla, i.e. a \sr--\ca--\sr three-ion chain; the IP-mode frequency is $2\pi\times730$~kHz.  The two \sr ions are entangled after a gate duration of 61~$\mu$s.  (d) Parity flopping curve (see text) for the \ca--\ca gate shown in part (a). (e) Parity flopping curve for the \sr--\sr gate shown in part (b).  (f) Parity flopping curve for the \sr--\ca--\sr gate shown in part (c).  Lines in (d), (e), and (f) are sinusoidal fits to the data with constrained periods; the best-fit amplitude is used to calculate the Bell-state fidelity.  The offset phases in these plots are related to $\phi$ and can be zeroed via adjustment of the bichromatic drive fields with respect to the carrier.}
\label{fig:same_species}
\end{figure*}

\section{Single-Species Two-Qubit Gates}

Single-species quantum logic gates are performed with either two {\ca} or two {\sr} ions in the trap such that the ions form a linear crystal oriented along the trap axis, with ions spaced a few microns apart.  After sideband cooling and state preparation as described above, the MS interaction is brought about by applying light detuned near the IP vibrational mode sidebands of the $|1\rangle \rightarrow |0\rangle$ transition; two frequencies are simultaneously applied to a single-pass AOM to produce a bichromatic light field with components detuned by $\delta$ above the blue sideband and below the red sideband of the mode $\beta$ to be driven.  This bichromatic field can be thought of as a drive resonant with the qubit carrier transition, but modulated at the beat frequency $\omega_{\beta}+\delta$. The bichromatic field is coupled into a single-mode, polarization-maintaining optical fiber and directed to the ions. Starting from $|11\rangle$, the joint qubit state is coherently driven between $|11\rangle$ and $|00\rangle$ via multiple pathways through $|10\rangle$ and $|01\rangle$ using the joint motional state.  If this evolution is stopped after a time $t_{g}$ as described above, the ions will be in a coherent superposition of $|11\rangle$ and $|00\rangle$, nominally the Bell state $|\Phi_{+\phi}\rangle = \frac{1}{\sqrt{2}}(|00\rangle + e^{i\phi}|11\rangle)$ (the value of $\phi$ is can be adjusted via the phases of the AOM RF drive signals used to create the bichromatic field; it must have a constant relative phase relationship with following analysis pulses).  We apply the bichromatic gate pulses with an additional asymmetric detuning, calibrated separately, to compensate for AC Stark shifts from other electronic levels, and we also shape the pulse amplitude in time to minimize dependence on the initial bichromatic phase~\cite{KirchmairHotGates2009}.

To characterize gate operation, we estimate the created Bell-state fidelity by measuring the four elements of the resulting two-qubit density matrix $\rho$ that would be nonzero in the case of ideal creation of $|\Phi_{+\phi}\rangle$.  The two diagonal elements are computed from the probabilities to measure $|00\rangle$ and $|11\rangle$ after the entangling gate operation, $P_{00}$ and $P_{11}$, respectively. The state population measurements are typically repeated thousands of times in order to precisely determine these values.  The off-diagonal elements are calculated using an auxiliary ``parity-flopping'' measurement in which a $\pi/2$-pulse around an axis in the equatorial plane of the Bloch sphere of varying phase angle $\chi$ with respect to the $X$ axis is applied uniformly to the ions in the created entangled state.  This experiment effectively rotates the coherences into the populations, and the parity of the populations, defined as $(P_{\chi,00}+P_{\chi,11})-(P_{\chi,01}+P_{\chi,10})$, will oscillate with a period of $\pi$ in $\chi$ for a two-qubit maximally entangled state.  The amplitude $C_{\textrm{PF}}$ of this oscillation gives a direct measure of the off-diagonal elements.  We calculate the Bell-state fidelity as $F_{|\Phi_{+\phi}\rangle}\equiv \langle \Phi_{+\phi}|\rho|\Phi_{+\phi} \rangle=(P_{00} + P_{11} + C_{\textrm{PF}})/2$.

Fig.~\ref{fig:same_species} shows results for {\ca} and {\sr} single-species, two-qubit gates; both the measured state populations as a function of gate-interaction duration and the parity-flopping curves are shown.  In the case of ideal evolution, the Bell state $|\Phi_{+\phi}\rangle$ is created at the second zero of the combined population of the $|10\rangle$ and $|01\rangle$ states (``1-bright'').  As can be seen, this time was (coincidentally) just over 70~$\mu$s in both cases; the achieved Bell-state error is 1.2(2)\% and 2.5(2)\% for \ca--{\ca} and \sr--\sr, respectively.  Leading error sources are believed to be due to state dephasing from cryocooler vibrations and laser phase fluctuations.  Cryocooler vibrations, which bring about an oscillation in the trap, and hence ion, location on the 10--100~nm scale with respect to the delivery optics can lead to shot-to-shot variations in the gate-laser phase at the ion location.  Laser instability is also a direct limit to optical qubit coherence time; we place an upper limit on the laser bandwidth of 50~Hz through single-ion Ramsey decay.  Sources of error which we expect to come in at a lower level include intensity fluctuations, due to power fluctuations and beam pointing instability, which lead to variations of the Rabi frequency and fluctuating AC Stark shifts due to additional levels in the ions' electronic structure.

We have also performed an entangling quantum gate between two computational ions in the presence of a third ion of a different species, a primitive described in Sec.~\ref{sec:arch}.  Our implementation consists of an MS gate performed between two {\sr} ions in a crystal with a \sr--\ca--{\sr} configuration (see Fig.~\ref{fig:nbrings}b).  Initialization for these experiments begins with Doppler cooling using the {\ca} ancilla and quenching of the $|0\rangle$ states of the computational {\sr} ions to prepare them in $|11\rangle$, followed by resolved sideband cooling on the IP vibrational mode using the ancilla only.  The gate is then performed with light resonant with the computational ions only, with the bichromatic field tuned near the IP mode, as in the \sr--{\sr} gate described above.  Fig.~\ref{fig:same_species} shows the gate flopping and parity flopping curves for this gate; we achieve a Bell state with infidelity of 0.043(3) in 61~$\mu$s.  Beyond the sources of imperfection mentioned above when discussing the \sr--{\sr} gate, additional error sources include mode coupling to the uncooled spectator modes, particularly the stretch mode in which the center {\ca} ion does not participate.

\section{Dual-Species Two-Qubit Gates}

\begin{figure}[t !]
\parbox[b][2.27 in][t]{0.01 in}{a)}\includegraphics[width = 0.8 \columnwidth]{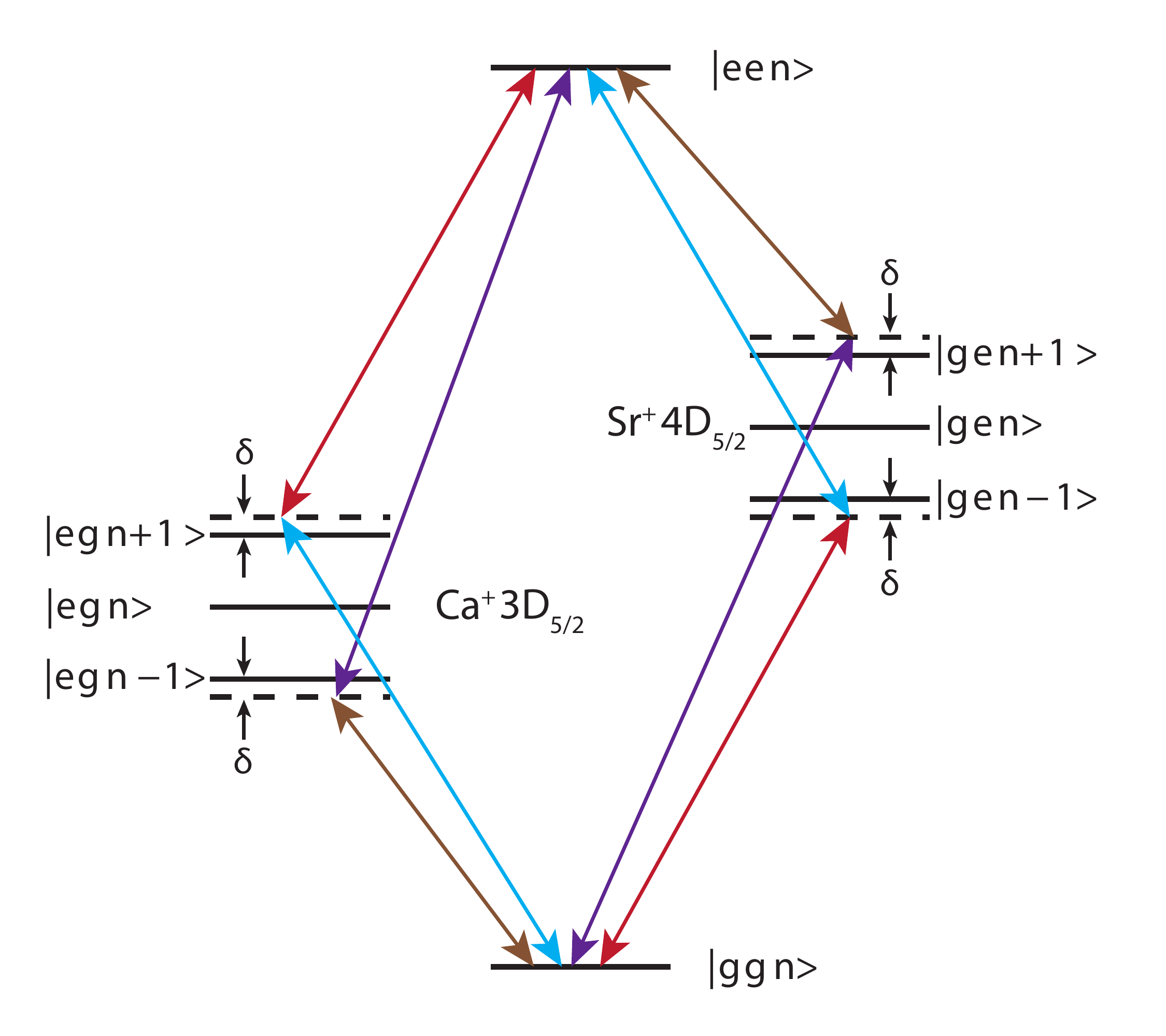} \\
\parbox[b][1.9 in][t]{0.2 in}{b)}\includegraphics[width = 0.75 \columnwidth]{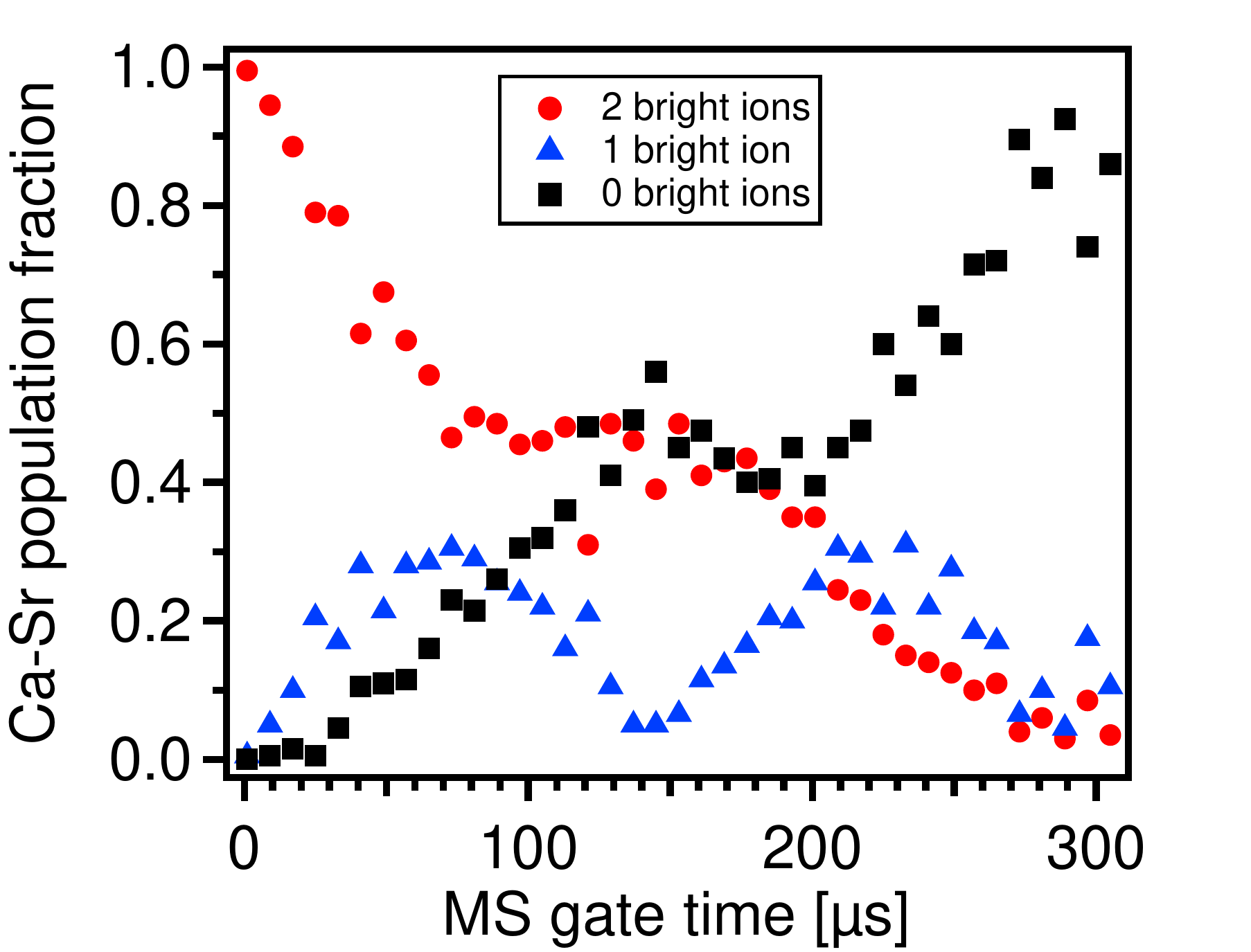} \\
\parbox[b][1.85 in][t]{0.2 in}{c)}\includegraphics[width = 0.75 \columnwidth]{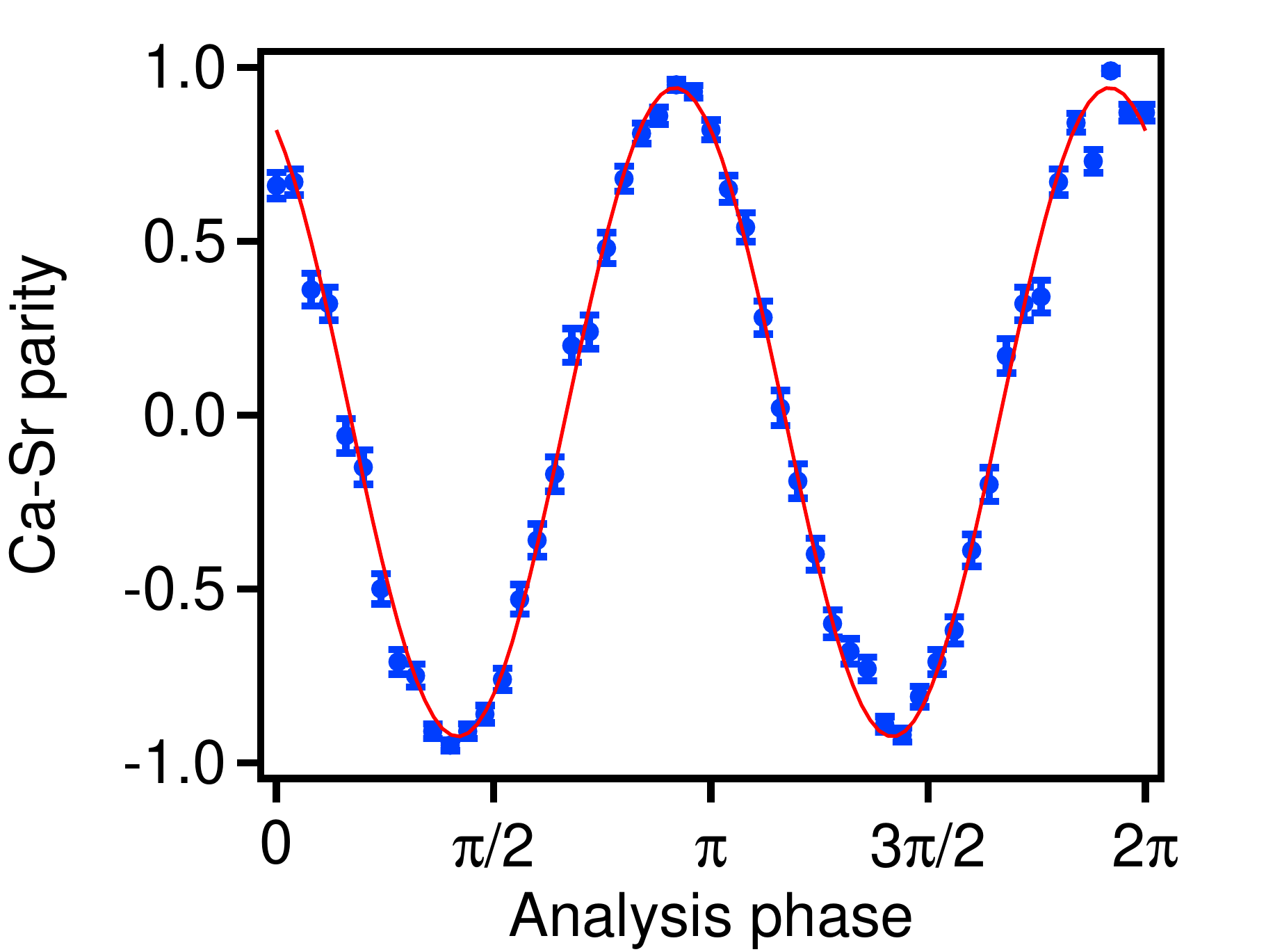}
\caption{Two-qubit entangling gate with ions of different species.  (a) Level structure used in the \ca--{\sr} M{\o}lmer-S{\o}rensen (MS) gate; The $|0\rangle,|1\rangle$ notation has been replaced with $|e\rangle,|g\rangle$ here to avoid confusion with the shared motional state occupation denoted by $n$. (b)  Representative measured state populations are plotted as a function of duration of the gate pulse, here for a gate performed on the in-phase mode at $2\pi\times770$~kHz.  A maximally entangled state is created here after 140~$\mu$s. The highest gate fidelities were achieved using a slightly lower Rabi frequency and correspondingly longer gate time of 160~$\mu$s. (c) Parity flopping for the 160~$\mu$s gate, where the line is a sinusoidal fit with constrained period.}
\label{fig:dual_species}
\end{figure}

We have also demonstrated a dual-species MS entangling operation.  To create the Bell state $|\Phi_{+\phi}\rangle$ between \ca and {\sr} in a two-ion crystal, we start by Doppler and resolved-sideband cooling both the axial IP and OOP modes using the {\sr} ion. Following optical pumping to bring the ions to the $|11\rangle$ state, we simultaneously apply two bichromatic light fields, one to couple the internal state of each species to the shared motion, each detuned from the IP vibrational mode (see Fig.~\ref{fig:dual_species}a) by~$\delta$.  The 674~nm and 729~nm beams are oriented parallel to the trap axis and anti-parallel to each other. As in the single-species case, we ramp up and down the pulse amplitude at the beginning and end of the interaction to avoid both dependence on the initial phase between the red- and blue-sideband drives and off-resonant excitation of the $|1\rangle \rightarrow |0\rangle$ carrier transition~\cite{KirchmairHotGates2009}.

An additional consideration with MS gates on two different optical transitions on two different ion species is the relative phase of the force on each ion; this relationship is dictated by the relative phase of the red- and blue-sideband phase differences on each bichromatic beam pair at the ions' location.  The distance between the ions (${\sim}3$~$\mu$m) and the difference in the optical path lengths between the two bichromatic pairs (${\sim}100$~mm) are both very small compared to the the distance between maxima of the amplitude-modulated waveform of each bichromatic pair (${\sim}100$~m, set by $\omega_{\beta}+\delta$, which is on the order of megahertz), so when the RF phases of the fields driving the AOMs producing the bichromatic fields are in phase, the force on the ions is as well.  The four AOM RF drive signals, two each driving the 674~nm and 729~nm AOMs, are all phase coherent, driven from the same clock, allowing control and maintenance of shot-to-shot phase coherence of the force on the two ions.  While the 674~nm and 729~nm lasers each need to be coherent over the course of each experiment such that the analysis pulse phase (and that of any subsequent algorithmic logic pulses) is coherent with the beams driving the gate, the relative optical phase between the two lasers does not need to be constant; only the relative phase between the red- and blue-sideband component phase {\it differences} must be maintained.

We extract the parity contrast by scanning the phases of $\pi/2$ analysis pulses for {\ca} and {\sr} applied simultaneously after the completion of the MS gate. In Fig.~\ref{fig:dual_species}b we plot a representative population-flopping curve (taken with a slightly higher Rabi frequency than that used to calculate the state fidelity) and in Fig.~\ref{fig:dual_species}c the parity-flopping curve for which the produced Bell state has a measured error of 0.057(3) and a duration of 160~$\mu$s. The mixed-species gate speed is limited here by the achievable sideband Rabi frequency for the {\ca} ion. The amplitude of the normalized eigenvector $\text{b}_{\mathrm{Ca, IP}}$ and the corresponding Lamb-Dicke parameter $\eta_{\mathrm{Ca, IP}}$ are significantly lower for the chosen in-phase mode than in the single-species chain (see Table~\ref{table:freqs} in the Appendix).  We expect some level of rejection of gate-laser-field noise common to both ions in same-species gates.  On the contrary, for dual-species gates, the ions are driven by different lasers.  Hence, we expect that effects such as differential phase and amplitude noise between the 674~nm and 729~nm light at the ion positions, leading to phase-space displacement variation and  additional fluctuating AC-Stark shifts, are the primary causes for the larger error in dual-species MS gates.



\section{Discussion}

\begin{table}[b]
\caption{Multi-qubit logic gates in the \ca/{\sr} system in this and previous work.  Gate type, gate duration, and infidelity are listed for M{\o}lmer-S{\o}rensen (MS) gates (this work) and quantum-logic spectroscopy (QLS) transfer operation from {\ca} to {\sr} (previous work~\cite{BruzewiczQLAR2017}). (*)~The gate was performed between the {\sr} ions with {\ca} used as a coolant ancilla.  ($\dagger$)~The QLS transfer requires two consecutive sideband pulses; the total time is given.  Also, QLS as demonstrated in~\cite{BruzewiczQLAR2017} maintains state populations only, hence here the inaccuracy (not infidelity) is reported.}

\begin{ruledtabular}
\begin{tabular}{lccc}

\multirow{2}{*}{Species} & \multirow{2}{*}{Gate type} & Gate time          &  Gate error \\
                         &              &  $t_{g}$ ($\mu$s)  &  $1-F_{|\Phi_{+i}\rangle}$ \\

\hline
\ca--\ \ca               &  MS     &   71          &  0.012(2)  \\
\sr--\ \sr               &  MS     &   72          &  0.025(2)  \\
\sr--\ \ca--\ \sr (*)    &  MS     &   61          &  0.043(3)  \\
\ca--\ \sr               &  MS     &   160         &  0.057(3)  \\
\ca--\ \sr (${\dagger}$) &  QLS     &   96          &  0.04(1)  \\

\end{tabular}
\end{ruledtabular}
\label{tab:fidelities}
\end{table}


The achieved infidelities and gate times for the \ca/{\sr} quantum logic primitives demonstrated here are listed in Table~\ref{tab:fidelities} along with the previously reported quantum-logic-assisted readout for the same two species.  The achieved error probabilities are not due to fundamental sources, and so we believe they can be reduced with technological improvements in qubit-laser frequency and amplitude stability at the ions' location.  This, along with the relative convenience of control methodologies for this pair of ion qubits, leads us to expect that these primitives will form the basis for more complex ancilla-assisted quantum information processing in larger \ca/{\sr} systems.

A notable aspect of multi-qubit operations in this particular system is the presence of optical-frequency qubits in both species, as we have demonstrated here.  The presence of metastable $D$ states allows for high-efficiency electron-shelving-based state detection, with the added potential for relatively lower optical power requirements for 10--100~$\mu$s two-qubit gate durations when compared to Raman-based gates, or very low ultimate error rates for direct optical single- and two-qubit operations~\cite{doi:10.1063/1.5088164}.  Additionally, {\ca} and {\sr} both possess optical qubits with qubit transition frequency in the red to near-infrared. This means that the control fields with the greatest requirements for optical power and frequency stability---those used for quantum gates---are at more technologically convenient wavelengths when compared to those used for Raman excitation, which are typically detuned from the higher-energy $S \rightarrow P$ transitions.  Optical elements such as crystals, fibers, and mirror coatings all perform better away from the ultraviolet. Furthermore, integrated waveguides and related photonics devices have lower loss at longer wavelengths~\cite{SoraceAgaskarSPIE2018}, which is of particular interest for applications benefiting from control of large arrays of ions.

A consideration when employing ancilla ions of a different species from the logic ions is vibrational mode structure.  Driving coherent-displacement-based gates near accidental near-degeneracies should be avoided for the highest-fidelity logic-gate operation, and in some cases species pairs may be chosen to avoid such coincidences.  This appears to be straightforward for typical gate durations, but the problem could become more pronounced for high-speed operations, where more precisely tailored amplitude-shaped pulses may be required to account for driving multiple modes~\cite{schafer2018fast}.

\section{Acknowledgments}

We thank Vladimir Bolkhovsky for trap fabrication, George Fitch for layout assistance, and Peter Murphy, Chris Thoummaraj, and Karen Magoon for assistance with chip packaging.  We are also grateful to Robert Niffenegger and Garrett Simon for comments on the manuscript.  This work was sponsored by the Under Secretary of Defense for Research and Engineering under Air Force contract number FA8721-05-C-0002. Opinions, interpretations, conclusions, and recommendations are those of the authors and are not necessarily endorsed by the United States Government.

\appendix*
\section{Mode Parameter Calculation for Dual-Species Ion Chains and Lamb-Dicke Parameters for Various {Ca}$^{+}$/{Sr}$^{+}$ Crystals}
\label{appendix1}

For the two-ion mixed-species chain, there are only two axial modes, and their frequency ratio is given by~\cite{WubennaSympCooling2012}:
\begin{equation}
\frac{\omega_{\mathrm{OOP}}}{\omega_{\mathrm{IP}}}=\sqrt{\frac{1+\mu+\sqrt{1-\mu+\mu^{2}}}{1+\mu-\sqrt{1-\mu+\mu^{2}}}},    
\end{equation}
where $\mu$ is the ion mass ratio and $\omega_{\mathrm{IP}}$ and $\omega_{\mathrm{OOP}}$ are the  frequencies of the in-phase and out-of-phase normal modes. The ion-dependent Lamb-Dicke parameter for ion $j$ with respect to the normal mode $\beta$ is given by~\cite{home2013quantum}: 
\begin{equation}
\eta_{j,\beta}=\sqrt{\frac{\hbar}{2m_{j}\omega_{\beta}}}\textbf{k}_{j}\cdot\textbf{b}_{j,\beta},
\label{eq:ldp}
\end{equation}
where $\textbf{k}_{j}$ is the wavevector of the qubit laser, $\textbf{b}_{j,\beta}$ is the normalized motional eigenvector for ion $j$ in normal mode $\beta$, $\omega_{\beta}$ is the oscillation frequency, $m_{j}$ is the ion mass, and $\hbar$ is the reduced Planck constant. The amplitudes of the motional eigenvectors $\text{b}_{j,\beta}$ can be calculated numerically for arbitrary chain lengths and configurations~\cite{home2013quantum}. In addition, simple closed form expressions exist for the two-ion chain~\cite{WubennaSympCooling2012}, to wit, for ions $i,j$:
\begin{align}
    \text{b}_{j,\mathrm{IP}}&=\sqrt{\frac{1-\tilde{\mu}+\sqrt{1-\tilde{\mu}+\tilde{\mu}^{2}}}{2\sqrt{1-\tilde{\mu}+\tilde{\mu}^{2}}}}, \nonumber\\
    \text{b}_{j,\mathrm{OOP}}^{2}&=1-\text{b}_{j,\mathrm{IP}}^{2},
\end{align}
where $\tilde{\mu}$ is the mass ratio expressed as $\tilde{\mu}=m_{i}/m_{j}$.

\begin{table*}
\begin{ruledtabular}
\caption{Calculated axial mode frequencies $\omega_{\beta}$, normalized mode eigenvector amplitudes $\text{b}_{j}$, and associated Lamb-Dicke parameters $\eta_{j}$ for mixed-species two-ion and symmetric three-ion chains made up of combinations of two different isotopes each of {\ca} and {\sr}. Frequencies are normalized by the single-ion frequency of the relevant {\sr} isotope in the third column and normalized to the lowest frequency mode ($\omega_{\mathrm{IP}}$) for each ion configuration to highlight possible higher order sideband degeneracies in the fourth column. To evaluate the Lamb-Dicke parameters, we calculate the necessary axial frequencies for  $\omega_{\mathrm{Sr}^{+}}=2\pi\times660$~kHz, the corresponding single-{\sr} trap frequency used for the dual-species gate demonstrated here. We also assume that the laser wavevector $\textbf{k}_{j}$ is parallel to the ion motion.}
\renewcommand{\arraystretch}{1}
\renewcommand\tabcolsep{6pt}
\begin{tabular}{l c c c c c c c}

Configuration& Axial mode~$\beta$ & $\omega_{\beta}/\omega_{\mathrm{Sr}^{+}}$ & $\omega_{\beta}/\omega_{\mathrm{IP}}$ & $\text{b}_{\mathrm{Sr}^{+}}$ &$\text{b}_{\mathrm{Ca}^{+}}$ & $\eta_{\mathrm{Sr}^{+}}$ & $\eta_{\mathrm{Ca}^{+}}$\\
\hline
\hline

\multirow{2}{*}{$^{40}\mathrm{Ca}^{+}\mathrm{-}^{88}\mathrm{Sr}^{+}$} & In-phase & 1.137 & 1 & 0.902 & 0.431 & $0.074$ & $0.048$ \\
             & Out-of-phase & 2.260 & 1.988 & 0.431 & 0.902 & $0.025$ & $0.072$  \\
\hline

\multirow{3}{*}{$^{40}\mathrm{Ca}^{+}\mathrm{-}^{88}\mathrm{Sr}^{+}\mathrm{-}^{40}\mathrm{Ca}^{+}$} & In-phase & 1.232 & 1 & 0.781 & 0.441 & $0.061$ & $0.047$\\
 & Stretch & 2.569  & 2.085 & 0 & 0.707 & 0 & $0.052$ \\
 & Alternating & 2.898 & 2.352 & -0.624 & 0.552 & $0.032$ & $0.039$\\
\hline
\multirow{3}{*}{$^{88}\mathrm{Sr}^{+}\mathrm{-}^{40}\mathrm{Ca}^{+}\mathrm{-}^{88}\mathrm{Sr}^{+}$} & In-phase & 1.095 & 1 & 0.652 & 0.385 & $0.054$ & $0.044$\\
 & Stretch & 1.732 & 1.582 & 0.707 & 0 & $0.047$ & 0 \\
 & Alternating & 3.262 & 2.979 & 0.272 & -0.923 & $0.013$ & $0.061$\\
\hline
\hline
\multirow{2}{*}{$^{43}\mathrm{Ca}^{+}\mathrm{-}^{88}\mathrm{Sr}^{+}$}
& In-phase & 1.129 & 1 & 0.892 & 0.453 & $0.073$ & $0.049$ \\
& Out-of-phase & 2.195 & 1.945 & 0.453 & 0.892 & $0.026$ & $0.069$  \\
\hline
\multirow{3}{*}{$^{43}\mathrm{Ca}^{+}\mathrm{-}^{88}\mathrm{Sr}^{+}\mathrm{-}^{43}\mathrm{Ca}^{+}$}
 & In-phase & 1.215 & 1 & 0.765 & 0.455 & $0.060$ & $0.048$\\
 & Stretch & 2.478  & 2.040 & 0 & 0.707 & 0 & $0.052$\\
 & Alternating & 2.836 & 2.335 & -0.644 & 0.541 & $0.033$ & $0.037$\\
 \hline
\multirow{3}{*}{$^{88}\mathrm{Sr}^{+}\mathrm{-}^{43}\mathrm{Ca}^{+}\mathrm{-}^{88}\mathrm{Sr}^{+}$}
 & In-phase & 1.089 & 1 & 0.648 & 0.400 & $0.054$ & $0.044$\\
 & Stretch & 1.732 & 1.590 & 0.707 & 0 & $0.047$ & 0  \\
 & Alternating & 3.164 & 2.905 & 0.283 & -0.916 & $0.014$ & $0.059$\\
 \hline
 \hline
\multirow{2}{*}{$^{40}\mathrm{Ca}^{+}\mathrm{-}^{86}\mathrm{Sr}^{+}$}
& In-phase & 1.134 & 1 & 0.899 & 0.438 & $0.074$ & $0.049$  \\
& Out-of-phase & 2.239 & 1.974 & 0.438 & 0.899 & $0.026$ & $0.072$  \\
\hline
\multirow{3}{*}{$^{40}\mathrm{Ca}^{+}\mathrm{-}^{86}\mathrm{Sr}^{+}\mathrm{-}^{40}\mathrm{Ca}^{+}$}
 & In-phase & 1.227 & 1 & 0.776 & 0.446 & $0.062$ & $0.048$\\
 & Stretch & 2.540  & 2.070 & 0 & 0.707 & 0 & $0.053$ \\
 & Alternating & 2.878 & 2.346 & -0.630 & 0.549 & $0.033$ & $0.038$\\
 \hline
\multirow{3}{*}{$^{86}\mathrm{Sr}^{+}\mathrm{-}^{40}\mathrm{Ca}^{+}\mathrm{-}^{86}\mathrm{Sr}^{+}$}
 & In-phase & 1.093 & 1 & 0.651 & 0.390 & $0.055$ & $0.044$\\
 & Stretch & 1.732 & 1.584 & 0.707 & 0 & $0.047$ & 0  \\
 & Alternating & 3.230 & 2.955 & 0.276 & -0.921 & $0.013$ & $0.061$\\
\hline 
\hline
\multirow{2}{*}{$^{43}\mathrm{Ca}^{+}\mathrm{-}^{86}\mathrm{Sr}^{+}$}
& In-phase & 1.126 & 1 & 0.888 & 0.460 & $0.074$ & $0.050$  \\
& Out-of-phase & 2.175 & 1.932 & 0.460 & 0.888 & $0.027$ & $0.069$   \\
\hline
\multirow{3}{*}{$^{43}\mathrm{Ca}^{+}\mathrm{-}^{86}\mathrm{Sr}^{+}\mathrm{-}^{43}\mathrm{Ca}^{+}$}
 & In-phase & 1.209 & 1 & 0.760 & 0.460 & $0.061$ & $0.048$ \\
 & Stretch & 2.449  & 2.026 & 0 & 0.707 & 0 & $0.052$ \\
 & Alternating & 2.818 & 2.331 & -0.650 & 0.537 & $0.034$ & $0.037$ \\
\hline
\multirow{3}{*}{$^{86}\mathrm{Sr}^{+}\mathrm{-}^{43}\mathrm{Ca}^{+}\mathrm{-}^{86}\mathrm{Sr}^{+}$}
 & In-phase & 1.087 & 1 & 0.646 & 0.405 & $0.054$ & $0.045$\\
 & Stretch & 1.732 & 1.594 & 0.707 & 0 & $0.047$ & 0  \\
 & Alternating & 3.133 & 2.883 & 0.286 & -0.914 & $0.014$ & $0.059$\\

\end{tabular}
\label{table:freqs}
\end{ruledtabular}
\end{table*}

Analogous calculations of the normal mode frequencies and Lamb-Dicke parameters in a symmetric three-ion chain, such as the \sr--\ca--{\sr} chain used here, can also be made.  The mode frequencies for a symmetric chain $i_{1}$--$j$--$i_{2}$ are given by~\cite{PhysRevA.61.032310} 
\begin{align}
\omega_{\mathrm{IP}}&=\sqrt{\frac{13}{10}+\frac{21-\sqrt{441-34\tilde{\mu}^{-1}+169\tilde{\mu}^{-2}}}{10\tilde{\mu}^{-1}}}\, \omega_{i}, \nonumber\\
\omega_{\mathrm{Stretch}}&=\sqrt{3}\, \omega_{i}, \\
\omega_{\mathrm{Alt}}&=\sqrt{\frac{13}{10}+\frac{21+\sqrt{441-34\tilde{\mu}^{-1}+169\tilde{\mu}^{-2}}}{10\tilde{\mu}^{-1}}}\, \omega_{i}, \nonumber
\end{align}
where $\omega_{i}$ is the axial frequency of a single ion with mass $m_{i}$. Here $\omega_{\mathrm{Stretch}}$ and $\omega_{\mathrm{Alt}}$ are mode frequencies for the second, breathing-type axial mode and the third, alternating-ion-motion axial mode, respectively (See Fig.~\ref{fig:nbrings}b).  The definition of $\tilde{\mu}$ used here is the inverse of $\mu$ used in Ref.~\cite{PhysRevA.61.032310} but has been chosen in order to be consistent with the convention used in the previous discussion of two-ion chains.

In contrast to the two-ion mixed-species chain, the normalized mode eigenvectors of a symmetric three-ion chain are dependent on the normal mode frequencies. These eigenvectors $\textbf{x}_{\beta}$ for ions $i_{1}$--$j$--$i_{2}$ in mode $\beta$ can be calculated as~\cite{PhysRevA.61.032310}
\begin{align}
    \textbf{x}_{\mathrm{IP}}&=N_{\mathrm{IP}}\bigg(1,\frac{13-5(\omega_{\mathrm{IP}}/\omega_{i})^{2}}{8\sqrt{\tilde{\mu}}},1\bigg), \\
    \textbf{x}_{\mathrm{Stretch}}&=N_{\mathrm{Stretch}}\bigg(1,0,-1\bigg), \\
    \textbf{x}_{\mathrm{Alt}}&=N_{\mathrm{Alt}}\bigg(1,\frac{13-5(\omega_{\mathrm{Alt}}/\omega_{i})^{2}}{8\sqrt{\tilde{\mu}}},1\bigg), 
\end{align}
where $\textbf{x}_{\beta}=(\text{b}_{i_{1},\beta},\text{b}_{j,\beta},\text{b}_{i_{2},\beta})$ and $N_{\beta}$ are normalization constants. The Lamb-Dicke parameters can then be calculated using Eq.~\ref{eq:ldp}.

In Table~\ref{table:freqs}, we list the mode parameters and Lamb-Dicke parameters (for a particular axial trap frequency) for dual-species, two- and three-ion crystals containing various common isotopes of {\ca} and \sr.  Besides allowing for the inclusion or exclusion of hyperfine structure in one or the other species, different isotopes lead to slightly different mode frequencies, providing the flexibility to avoid accidental overlaps between the target mode sidebands and higher-order sidebands of spectator modes.  Small changes in mode frequencies can make a large difference as  the relevant comparison is the frequency shift to the detuning from the sidebands, typically several kilohertz to a few tens of kilohertz (see main text).

\bibliography{gate_paper_v1}

\end{document}